# Data visualization in political and social sciences

Andrei Zinovyev

Institut Curie, Paris, France
zinovyev@gmail.com

The basic **objective of data visualization** is **to provide an efficient graphical display for summarizing and reasoning about quantitative information**. Data visualization should be distinguished from other types of visualization used in political science (more general information and knowledge visualization, concept visualization, strategy and workflow visualization, metaphor visualization, etc.) and is more specific to representing quantitative data existing in the form of *numerical tables*.

During the last decades, political science has accumulated a large corpus of various kinds of data such as comprehensive factbooks and atlases, characterizing all or most of existing states by multiple and objectively assessed numerical indicators within certain time lapse (examples are *OECD Factbook* and *Political Atlas of the Modern World* projects). As a consequence, there exists a continuous trend for political science to gradually become a more quantitative scientific field and to use quantitative information in the analysis and reasoning. It is believed that any objective analysis in political science must be multidimensional and combine various sources of quantitative information; however, human capabilities for perception of large massifs of numerical information are limited. Hence, methods and approaches for visualization of quantitative and qualitative data (and, especially *multivariate data*) is an extremely important topic in political science. Data visualization approaches can be classified into several groups, starting from creating informative charts and diagrams (statistical graphics and infographics) and ending with advanced statistical methods for visualizing multidimensional tables containing both quantitative and qualitative information.

Data visualization in political science takes advantage of recent developments in computer science and computer graphics, statistical methods, methods of information visualization, visual design and psychology. Data visualization in political science has certain specificity such as the frequent use of geographical maps, which creates a link with the well-developed field of geographical information systems (GIS). Also numerical tables in political science are frequently incomplete which makes important use of the methods allowing dealing with missing or uncertainly measured data entries.

There are two main types of numerical tables that can be a subject of data visualization. The first one is called *"object-feature" table*, where every row represents an observation or an object and every column correspond to a numerical feature or indicator commonly measured for the whole set of objects. An example of such an "object-feature" table is a factbook for a set of countries (see the Figure), where the objects are countries and the features are numerical indicators such as GDP per capita, employment rate, life expectancy, etc. The second type of numerical tables is called *connection or distance tables* where both rows and columns correspond to objects and at the intersection of a row and a column a numerical value is found characterizing a link between two objects. Alternatively, such tables can be represented in a form of *a list of links*. A typical example of a connection table is the table representing the migration rates or the mutual volumes of export and import of goods for a set of countries (at the intersection of the row A and the column B, the volume of export from the country A to the county B is found).

Data visualization plays several important roles in science: It 1) helps create informative illustrations of the data, recapitulating large amount of quantitative information on a diagram; 2) helps formulate new or supporting existing hypotheses from quantitative data; 3) guides a statistical analysis of data and checks its validity. One can also mention the role of infographics in creating images with a clear and visual message, based on numbers; thus, data visualization can serve as a powerful propagandistic or educational tool. Graphical display allows not only visualizing and analyzing the message contained in data but also remembering it, since for most of people, visual memory is more persistent than verbal or auditory memory (the phenomenon of *pictorial superiority*).

Several groups of methods for data visualization can be distinguished in political science:

- *Statistical graphics and infographics* with extensive use of color, form, size, shape and style to superimpose many quantitative variables in the same chart or diagram

- *Geographical information systems (GIS)* to visualize geographically-linked data

- *Graph visualization* or *network maps* for representing relations between objects

- Projection of multidimensional data on low-dimensional screens with further visualization, *data cartography*

Further in the article, a description of these principal groups of data visualization methods together with references to the examples of their application in political science is provided.

## Statistical graphics and infographics

The statistical graphics started to be used in science in the XVIII century but began to be widely exploited only from the end of XIX century. Many of the currently used types of charts were introduced by William Playfair (1759-1823), a Scottish political economist, and Johann Heinrich Lambert (1728-1777), a Swiss-German mathematician. With the appearance of computer-based technologies in the end of the XX century, the use of statistical graphics exploded and included the use of three-dimensional, dynamical and interactive data representations.

Lengler and Eppler suggested a "periodic table" of visualization methods with many examples of elements of statistical graphics. The most commonly used quantitative data displays can be classified into

- *Univariate plots*, designed to visualize value distribution of a single variable. The basic ones are *pie charts*, *bar charts*, *histograms*, *boxplots*.

- *Bivariate plots,* designed to visualize relations between two variables. These include various types of *scatterplots* and *line plots*. A particular important case is the line plot visualizing *time series* on which one variable is *time*.

- *Multivariate plots,* designed to visualize the values of several variables at the same time and allowing comparing them. Examples of these are *area charts*, *radar charts*, *mosaic plots*, *parallel coordinates plots* and more exquisite *Chernoff faces plots*.

This classification is elementary; however, modern statistical graphics usually superimpose several types of plots and use color, shape, size and style of their elements to increase the dimensionality of the plot. New types of data displays are invented continuously and proposed as a part of commonly used software for data analysis. As an example, the data table itself can serve as a data display (see the Figure). Moreover, the table can be converted into the more elaborated *information lense* display. Time series representation can be significantly enhanced by colorful *horizon graphs* that allow visualizing simultaneously hundreds of time series and identifying patterns in their behavior visually. *Treemaps* are a space-filling approach to showing hierarchies in which the rectangular screen space is divided into regions, and then each region is divided again for each level in the hierarchy (treemaps were used, for example, for visualization of the US budget). Intersections between sets of objects can be represented by *Venn diagrams*. Many useful graphical displays result from application of multidimensional statistical analysis techniques such as *factor analysis* (including *principal components analysis*) and *clustering* (see the Figure and the Multidimensional Data Cartography section below).

In the physical, biological and social sciences, scatterplot is the predominant type of data display (accordingly to Tufte, approximately 75% of graphs used in science are scatterplots). Scatterplot serves for two basic purposes: 1) to visually detect linear and non-linear relations between two variables; the human eye can do it efficiently and it is extremely robust to the effects of anomalous observations (*outliers*) and other aberrations in the data; and 2) to create a *map of data* allowing to visualize other information on top of it. In the case of a large number of points, the scatterplot can be improved by visualizing the *point density*, using *gradient shading* (see the Figure) *or isodensity levels*.

On the other hand, it was shown that the use of statistical graphics in news publications is mainly limited to univarite plots (if time series plots are not considered), mainly, barcharts and piecharts. Curiously, several Japanese news journals and the British *Economist* create notable exceptions and provide up to 9% of data visualizations in the form of bivariate plots.

One of the first books on the history of data visualization, "Graphic presentation", was written by Willard Brinton in 1939. It contains many examples of thoughtful data visualization in the pre-computer era.

## Geographical information systems

In 1854, John Snow depicted a cholera outbreak in London and detected the source of the epidemia using points to localize some individual cases (Tufte, 2001). This was one of the earliest applications of the *geographic method*, i.e. combining the use of a geographical map and statistical data. Nowadays, the use of geographical maps with several layers of information on top (also called *thematic maps* or *semantic layers*) is facilitated by Geographical Information Systems (GIS). GIS is the merging of *cartography*, *statistical analysis*, and *database technology*. Geovisualization tools have been used extensively in electoral studies, urbanization studies (LeGate, 2005), as well as in various geopolitical studies of empires, wars, boundaries and trade routes (Ward, 2002).

In the mid of 2000s several publicly available GIS systems appeared that are called *virtual globes*, with *Google Earth* being the most famous one. These systems allow the visualization of geographical maps and various types of global and local *semantic layers* mapped onto them in a highly dynamic, interactive fashion. Importantly, these systems themselves can collect information from users (*Public Pariticipation GIS*). Because of their wide use by the public for various purposes, these systems have high potential in being utilized in political and social studies and make a subject of several educational courses in social science (LeGates, 2005).

## Graph (network) visualization

Graph visualization (where graph is understood as a set of *vertices* connected by *arrows, directed or undirected*) also known as *network maps* is a graphical data display which allows to visualize a specific type of *connection tables* (see their definition above). On such a display, a vertex represents an object or an observation, and an arrow between two vertices represents a connection between two objects. Usually, only arrows representing non-empty connections (or having a strength exceeding a certain threshold) are introduced into the graph. On the network maps one can use the size and the color of vertices to visualize some features of the objects, and the thickness and the color of arrows to represent some features of connections between objects (typically, the strength of the connection). If the connections are asymmetrical (such as export and import flows, migration rates between countries) then two directed arrows connecting the same objects can be used to represent the asymmetric flow in both directions. Network maps can also represent distances between objects in the multidimensional space of multivariate observations.

Network maps are widely used for visualizing social network structures, trading relations between countries, migration rates, etc. A good example of such visualization is the *Mapping Globalization Project* of Princeton University.

The principal difficulty in using network map displays is that they easily become too entangled with a growing number of arrows. This poses technical challenges for 1) finding a good layout for placing vertices on a two-dimensional plane in order to minimize the number of arrow intersections or better visualize the internal structure of the graph; 2) creating handy browsers for huge network presentations. For example, in the *Walrus* network visualization tool, three-dimensional interactive network map representations are combined with *hyperbolic viewer* which creates a fisheye-like distortion in order to magnify (to zoom) a particular part of the graph while the rest of it remains less detailed and serves rather for orientation and navigation.

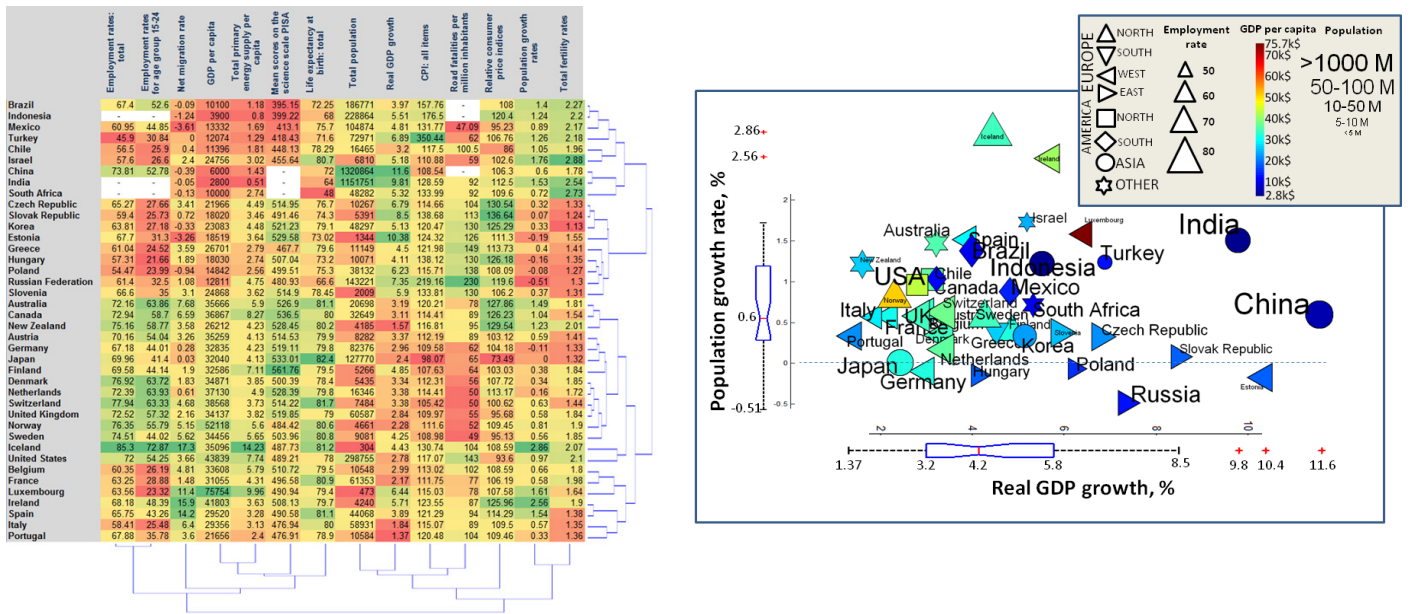

A)
B)

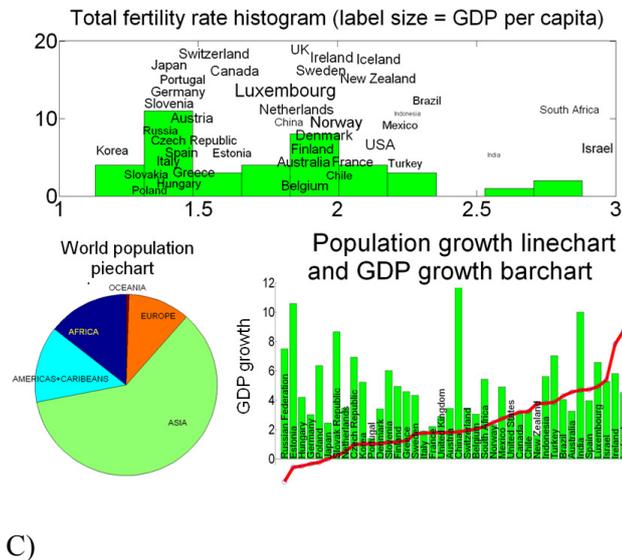
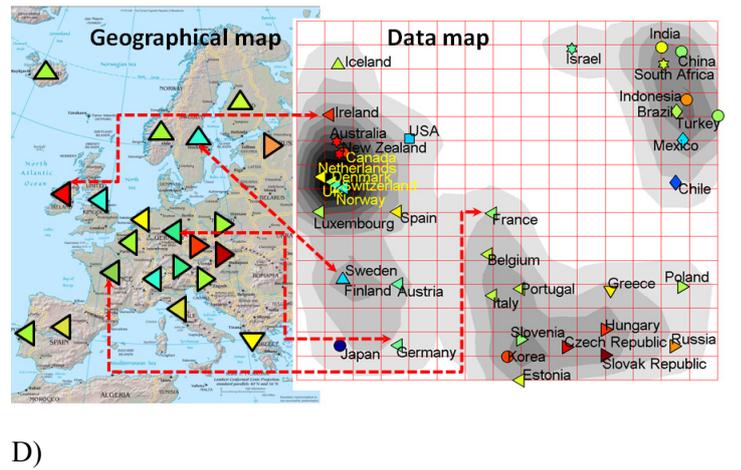

C)
D)

**Figure**. Examples of data visualization for a small subset of data (14 socio-economical numerical indicators for 40 countries) taken from OECD Factbook (for the year 2006). A) The table itself as a *graphical information display*. The green-red color map shows the value of the indicator (green – above the average, red – below the average, yellow – near the average). The rows and columns are reordered such that correlated indicators and statistical country profiles are ranked closely. The dendrogram on the right (or at the bottom) shows separation of countries into groups (or grouping correlated indicators) as suggested by *hierarchical clustering*. B) *Scatterplot* as a mean to visualize simultaneously values of 5 indicators (GDP and population growth rates using *abscissa and ordinate point position*, employment rate using *marker size*, GDP per capita using *marker color* and population using *text size of the data labels*). In addition, the geographical localization is visualized using the *form of the marker*. The boxplots on the axes depict the *median value*, lower and upper *quartiles*, extent of the data and *outlier values* (by red crosses). C) Examples of standard statistical graphs. D) *Data cartography* for multivariate data visualization. Non-linear data map is shown on the right and created with the use of elastic map algorithm implemented in ViDaExpert software (http://bioinfo.curie.fr/projects/vidaexpert). The gray shading represents the density of points (from this *semantic map*, four groups of countries can be distinguished). Geographical map on the left shows geographical localization of some of the points corresponding to European countries. Marker color here represents the relative consumer price index (from blue to red). The countries close on the geographical map are co-localized physically while the countries close on the data map are neighbors in the multidimensional space of indicators and have similar statistical profiles. The axes of the data map themselves do not have explicit interpretation; what matters is the *distance between points*, showing similar statistical profiles for points with a short distance and different ones for those with a bigger distance.

## Multidimensional data cartography

Inventors of statistical graphics thought of data displays as analogues of geographical maps. For bivariate data visualization plots, it was a crucial invention to replace geographical latitude and longitude by arbitrary measurement axes, which required the notion of a coordinate system formalized in the Descarte's *La Geometrie* written in 1637. Thus, on a scatterplot, instead of putting close geographically co-located objects, one puts close the points corresponding to similar combinations of measured $x$ and $y$ values.

For multivariate observations, one can formulate similar problem statement: how to map a set of objects (vectors) from a multidimensional space onto a two-dimensional plane (or into a three-dimensional space) such that the objects with similar numerical feature profiles would be located close and dissimilar objects would be located at a larger distance after projection. There are numbers of statistical methods aiming at producing such *data maps*. That is a part of *data exploratory analysis* field (see *Data Analysis, Exploratory*).

The most fundamental method for mapping multidimensional data into low-dimensional space is the *Principal Component Analysis* (PCA), invented by Karl Pearson (1857-1936). The method can be applied for the numerical tables of "object-feature" type. The set of table rows is represented as a cloud of data points (vectors) in the multidimensional space of features. Then PCA constructs an *optimal linear two-dimensional screen* (a plane) embedded into the multidimensional space of data such that the sum of squared distances between data points and the screen is minimal. After that, the data points can be projected onto the screen and the distribution of the projections is represented using standard statistical graphics techniques. The great advantage of such visualization is that it takes into account all numerical dimensions at the same time and not only two dimensions as on the standard bivariate scatterplots. The disadvantage of all multidimensional data mapping methods is that the new axes of the resulting scatterplot do not have explicit meaning (in general, they are complex functions of the initial numerical features). What matters on such data visualization displays are the relative distances between projections that represent the distances between objects in the initial multidimensional space. Method of principal components as a multivariate data visualization method was used in several studies, including Political Atlas of the Modern World (Melville et al, 2010). The classical PCA method can have non-linear generalizations (*principal manifolds* approach), and in this case it becomes more precise and informative in visualizing the data structure. In this case, a non-linear (curved) two-dimensional screen is constructed in the multidimensional space and used in the same way for projecting the data as in classical PCA (Gorban et al, 2008). Other multivariate data mapping techniques such as *Correspondence analysis*, *self-organizing maps*, *metro maps visualization of principal trees*, *multidimensional scaling*, *locally linear embedding*, *ISOMAP* have been recently developed and applied in political and social science.

Visualizing multivariate data by projection onto a low-dimensional screen is a subject of *multidimensional data cartography*: the projection creates a *data map* which is an alternative to the *geographical map* (see the Figure), and on which the objects (not necessarily linked to geography) with similar feature profiles are co-localized. For these data maps the methods of data visualization developed for GIS can be reutilized. For example, an atlas of semantic layers can be created, each layer corresponding to the values of a particular numerical attribute (see an example of such a system developed for the Political Atlas of the World at http://www.ihes.fr/~zinovyev/atlas).

## Data visualization problems and risks

Despite the undeniable role of data visualization in providing an efficient tool for reasoning on quantitative data, there are numbers of problems connected to possible misuse of graphical data displays. Potentially this can lead to a wrong interpretation of the message contained in the diagram.

Tufte and Wainer in their books provided numerous examples of mistakes or intentionally introduced distortions in a data visualization plot that can result in a message that is significantly different or even opposite to the one contained in the data. Most often these problems are related to misuse of axis scales, color palette or elements of design.

Bresciani and Eppler provided a two-dimensional classification of data visualization problems. First, the problems can be induced by *the designer* (intentionally or unintentionally) or by *the user* of the diagram. Second, these problems can be classified into *cognitive*, *emotional* and *social* ones. Cognitive problems can be connected to inappropriate use of graphical elements, lack of clarity, over-simplification or over-complexification of the graphical display, or induced by heterogeneity of target user groups (for example, women have a more accurate perception of the color palette than men). Emotional problems can be connected to a repellent content of graphical design. Social problems can be connected to cross-cultural differences of users (for example, in some eastern countries time is shown from right to left and the meaning of red and green is not identical to that accepted by western countries).

In the field of data cartography, the main problem of data visualization is the possible distortion of mutual distances when projecting data points from multidimensional to a low-dimensional space. Points distant in the multidimensional space can be projected at a short distance in the classical PCA method, and points close in the multidimensional space can be projected at a large distance in applications of non-linear data mapping techniques. There are visualization methods warning the user of the plot about possible distortions; however, in general, any conclusions derived from analyzing a data map should be verified carefully by rigorous statistical techniques of hypotheses testing.

Yet another source of problems in data visualization comes from the use of categorical or qualitative measurements for which no standardized and well-established graphical displays exist.

Data visualization risks should be distinguished from challenges posed to scientific data visualization in various fields. These challenges aim at making data visualization more informative and taking advantage of recent achievements in computer graphics, psychology, and computer science. For example, *ten top data visualization problems* were formulated, among which *ameliorating usability* and *scalability* of graphical displays as well as *shifting the visualization focus from visualizing static structures to visualizing dynamics*. Several international conferences (such as *IEEE Visualization* and *IEEE Information Vizualization*) provide a wide forum for answering these challenges by international and interdisciplinary research community effort.

Few selected websites devoted to data visualization and its application in sociology, economics and political science:

1. http://datavisualization.ch: The premier news and knowledge resource for data visualization and infographics.

2. http://www.gapminder.org/: An excellent dynamic and interactive data visualization tool and a database of socio-economical indicators collected for 200 years of the world history.

3. http://www.google.com/publicdata/: Google Public Data Explorer makes large datasets easy to explore, visualize and communicate. The public datasets available contain several factbooks (such as OECD factbook) that can be used in political science studies.

4. http://hdr.undp.org/en/reports/: Human Development Reports contain tons of examples of thoughtful data visualization, especially in recent reports.

5. http://qed.princeton.edu/main/MG/Data_and_Analysis: Mapping Globalization Project of Princeton University. The aim of the project is to visualize trading relations between countries.

## Further reading

1. Bresciani, S., Eppler, M.J. (2008). The risks of visualization: a classification of disadvantages associated with graphic representation of information. *ICA working paper* #1/2008.

2. Brinton, W. (1939). Graphic presentation. New York, Brinton associates. The book is available on-line at